\RequirePackage{fix-cm}
\documentclass[smallextended]{svjour3}       
\smartqed  
\usepackage{amssymb}
\usepackage{amsmath} 
\usepackage{graphics}
\usepackage{epsfig}
\usepackage{graphicx, subfigure}

 \usepackage{natbib}

\usepackage{color}
\usepackage{booktabs}
\usepackage{bbm}
\usepackage{caption}

\usepackage{floatrow}
\newfloatcommand{capbtabbox}{table}[][\FBwidth]
\usepackage{blindtext}

\newcommand{\vc}{{\bf c}}

\newcommand{\vx}{{\bf x}}

\newcommand{\vX}{{\bf X}}

\newcommand{\bqa}{\begin{eqnarray*}}
\newcommand{\eqa}{\end{eqnarray*}}
\newcommand{\bqan}{\begin{eqnarray}}
\newcommand{\eqan}{\end{eqnarray}}
\newcommand{\bit}{\begin{itemize}}
\newcommand{\eit}{\end{itemize}}
\newcommand{\ben}{\begin{enumerate}}
\newcommand{\een}{\end{enumerate}}
\newcommand{\beq}{\begin{equation}}
\newcommand{\eeq}{\end{equation}}
\newcommand{\bdes}{\begin{description}}
\newcommand{\edes}{\end{description}}
\begin{document}

\title{{A Basis Approach to Surface Clustering}
}


\author{Adriano Z. Zambom         \and
        Qing Wang \and
        Ronaldo Dias
}


\institute{Adriano Z. Zambom \at
              Department of Mathematics, California State University Northridge, USA \\
              Tel.: (818)677-2701\\
              \email{adriano.zambom@csun.edu}           
           \and
           Qing Wang \at
              Department of Mathematics, Wellesley College, USA
             \and
             Ronaldo Dias \at
             Department of Statistics, State University of Campinas, Brazil
}

\date{Received: date / Accepted: date}

\maketitle

\begin{abstract}
This paper presents a novel method for clustering surfaces. The proposal involves first using basis functions in a tensor product to smooth the data and thus reduce the dimension to a finite number of coefficients, and then using these estimated coefficients to cluster the surfaces via the $k$-means algorithm. An extension of the algorithm to clustering tensors is also discussed. We show that the proposed algorithm exhibits the property of strong consistency, with or without measurement errors, in correctly clustering the data as the sample size increases. Simulation studies suggest that the proposed method outperforms the benchmark $k$-means algorithm which uses the original vectorized data. In addition, an EGG real data example is considered to illustrate the practical application of the proposal.
\end{abstract}

\section{Introduction}

\indent \indent

Clustering objects in an infinite dimensional space is a challenging task given the complex nature of the data.
Although most data on a continuous domain are observed at a finite set or grid,  the computational cost may be too high or the direct application of a clustering procedure to the raw data may fail to capture the intrinsic stochasticity of the observations. Examples of such data structures with infinite dimensions include curves, surfaces, and tensors, which are in reality usually observed with errors. 
The main goal of this paper is to develop a novel clustering procedure for data sets whose elements are surfaces such as bivariate densities.
The idea is to first find an approximation of each surface by estimating the matrix (or tensor) of coefficients of a model in a finite dimensional space, thereby lowering the complexity of the data, and use these coefficients as the new data for a certain clustering method.

Since the seminal paper of \cite{Ward1963} introducing hierarchical clustering and the paper of \cite{Hartigan1975} and \cite{HartiganWong1979} discussing $k$-means clustering, developments and adaptations of these classical algorithms have been seen in a wide range of applications, such as in bioinformatics (\cite{Abu-Jamous15}, \cite{Duran19}), clinical psychiatry (\cite{Murcia19}), environmental policy (\cite{Hu19}), market segmentation (\cite{Wedel99}),  medicine (\cite{Udler18}), text mining (\cite{Abualigah18}), supply management (\cite{Blackhurst18}), and many other areas. 
The overall goal of these algorithms is to find partitions of the data based on distance metrics between elements. 
For instance, in (agglomerative) hierarchical clustering one produces a sequence of $n-1$ partitions of the data, starting with $n$ singleton clusters, and then merging the closest clusters together step by step until a single cluster of $n$ units is formed. The iterative $k$-means clustering algorithm starts with a set of $k$ initial cluster centers given as input based on a starting-point algorithm or previous knowledge. Each element of the data is then assigned a cluster membership in a way that the within-cluster sum of squares is minimized.

Clustering methods for curves, i.e. functional data clustering, have been explored by several researchers in the past few years.
\cite{IevaEtAl2013}, for example, performs clustering for multivariate functional data with a modification of the $k$-means clustering procedure. In its motivating example, the goal is to find  clusters of similar reconstructed and registered ECGs based on their functional forms and first derivatives. For the multivariate functional data, $\vX(t) = (X_1(t), \ldots, X_p(t))$ $(p\in\mathbb{Z}^+)$, where $t$  is in a compact subspace of $\mathbb{R}$ (often representing time),
\cite{MartinoEtAl2019} generalize the Mahalanobis distance in Hilbert spaces to create a measure of the distance between functional curves and use it to build a $k$-means clustering algorithm. Their setting  is different from our proposed method in that the time $t$ in \cite{MartinoEtAl2019} is the same for all components of the multivariate functional data, while in this paper we allow bivariate functions, such as $X(t_1,t_2)$, for example.
A non-exaustive list of recent literature that has studied functional data clustering includes \cite{AbrahamEtAl2003}, \cite{TokushigeEtAl2007}, \cite{YamamotoTerada2014}, \cite{WangEtAl2014}, \cite{GarciaEtAl2015}, \cite{BandeFuente2012}, \cite{TarpeyKinateder2003}, \cite{Floriello2011}, \cite{Yamamoto2012}, \cite{FerratyVieu2006}, and \cite{BoulleEtAl2014}.

In this paper we are interested in the generalization of clustering methods, such as the $k$-means algorithm, to surfaces and tensors. We propose using  basis functions in a tensor product as an approximation of the observed data, and then applying the estimated  coefficients of the basis functions to cluster the surfaces (or tensors) with the $k$-means algorithm. Simulations show that our proposed method significantly improves the accuracy of clustering compared to the baseline $k$-means algorithm applied directly to the raw vectorized data.

The remainder of the paper is organized as follows. In Section \ref{sec.method} we describe {the estimation procedure of the surfaces and the algorithm for clustering surfaces. Section \ref{sec.theory} shows some asymptotic results on the strong consistency of the algorithm in correctly clustering the data as the sample size increases. A generalization of this method to tensor products of higher dimensions is discussed in Section \ref{sec.tensor}. In Section \ref{sec.simul} we present simulations that assess the finite sample performance of the proposed method in comparison with the benchmark $k$-means.

\section{Methodology} \label{sec.method}

Let $\mathcal{S}^i := \mathcal{S}^i(x,y), (x,y) \in \mathcal{Q}$ be the underlying data generating process of the $i$th surface $(1\leq i\leq n)$, where $\mathcal{Q}$ is a compact subset of $\mathbb{R}^2$. Since data are in general discretely recorded and frequently contaminated with measurement errors, denote
\bqan\label{model}
z_j^i = \mathcal{S}^i(x_j^i, y_j^i)+\epsilon_j^i,\quad \ (1\leq j\leq m_i;1\leq i\leq n)
\eqan
 as the $m_i$ observed values of the $i$th surface at coordinates $(x_j^i, y_j^i)$, where $\epsilon_j^i$ is the measurement error which is assumed to be i.i.d. with mean 0 and constant finite variance $\sigma^2$.

Among several possible ways of representing functions and surfaces using basis functions such as wavelets (\cite{Mallat2008}), spline wavelets (\cite{Unser1997}), logsplines (\cite{KooperbergStone1992}), Fourier series, radial basis, in this paper we focus on B-splines (\cite{Boor1977}). The theoretical results we establish next are also valid for the aforementioned dimension reduction methods.
Assume that the surface $\mathcal{S}^i(x,y)$ can be well approximated by the smooth B-Spline tensor product,  defined as
\bqa
s^i(x,y,\boldsymbol{\Theta})=\sum_{r=1}^R\sum_{l=1}^L B_{x,r}(x)B_{y,l}(y)\theta_{rl}^i,
\eqa
where $\theta_{rl}^i$ are coefficients to be estimated, and $B_{x,r}(\cdot)$ and $B_{y,l}(\cdot)$ are B-spline basis functions that generate the spline spaces $\mathbb{S}_1 = \text{span}\{B_{x,1}, \ldots, B_{x,R}\}$ and $\mathbb{S}_2 = \text{span}\{B_{y,1}, \ldots, B_{y,L}\}$ respectively. $B_{x,r}(\cdot)$ and $B_{y,l}(\cdot)$ are polynomials of degree $p$ and $q$ respectively, so that $\mathbb{S}_1$ and $\mathbb{S}_2$ are piece-wise polynomials with $p-1$ and $q-1$ continuous derivatives. For ease of notation, we use the same degree and the same vector of knots for the B-spline basis functions for all the $n$ surfaces (\cite{deBoor1971}, \cite{deBoor1972}). In this paper we assume $R$ and $L$ are fixed, however, there are several methods in the literature that describe automatic procedures to obtain these values, see for example \cite{Unser1997} and \cite{DiasGamerman2002}.
The linear space of functions formed by the product of these two spaces is denoted by  $\mathbb{S}_1\otimes\mathbb{S}_2$. 
Because the Sobolev  space $\mathcal{H}^2 := \left\{ f:\int f^2 + \int(f')^2 + \int (f''^2) < \infty\right\}$ can be well approximated by $\mathbb{S}_1$ or $\mathbb{S}_2$ in their respective domains (\cite{karlin1973}, \cite{Reif1997}, \cite{LachoutEtAl2005}, \cite{LindemannLaValle2009}),
 the space  of smooth surfaces in $\mathcal{H}^2\otimes\mathcal{H}^2$ can be well approximated by $\mathbb{S}_1\otimes\mathbb{S}_2$.

The $R\times L$ matrix of coefficients $\boldsymbol{\Theta}^i=\{\theta_{rl}^i\}_{1\leq r\leq R;1\leq l\leq L}$ for each surface $i\ (i = 1, \ldots, n)$ observed with measurement error as specified in model (\ref{model}) can be estimated by minimizing the least squares errors
\begin{eqnarray*}
\text{vec}(\widehat{\boldsymbol{\Theta}}^i) &=& \arg\min_{\boldsymbol{\Theta}}\sum_{j=1}^{m_i}[z_j^i-s^i(x_j^i,y_j^i,\boldsymbol{\Theta})]^2\\
&=&\arg\min_{\boldsymbol{\Theta}}\sum_{j=1}^{m_i}[z_j^i-{\boldsymbol{B}^i_x(x_j^i)}^T\boldsymbol{\Theta}\boldsymbol{B}^i_y(y_j^i)]^2\\
&=&\arg\min_{\boldsymbol{\Theta}}\sum_{j=1}^{m_i}[z_j^i-({\boldsymbol{B}^i_y(y_j^i)}\otimes{\boldsymbol{B}^i_x(x_j^i)})^T\text{vec}(\boldsymbol{\Theta})]^2\\
&=&({M^i}^TM^i)^{-1}{M^i}^T\mathbf{z}^i
\end{eqnarray*}
where $\text{vec}(\boldsymbol{\Theta}^i)$ is the vectorization of the matrix $\boldsymbol{\Theta}^i$ arranged by columns,   $\mathbf{z}^i=(z_1^i,\ldots,z_{m_i}^i)^T$, ${\boldsymbol{B}^i_x}(x_j^i)=(B_1^i(x_j^i),\ldots,B_R^i(x_j^i))^T$, ${\boldsymbol{B}^i_y}(y_j^i)=(B_1^i(y_j^i),\ldots,B_L^i(y_j^i))^T$, and $M^i$ is the $m_i\times RL$ matrix with its $j$th row equal to the $1\times RL$ vector of $(\boldsymbol{B}^i_y(y_j^i)\otimes \boldsymbol{B}^i_x(x_j^i))^T$.

Surface $\mathcal{S}^i$ is hence summarized by the estimated matrix of parameters $\widehat{\boldsymbol{\Theta}}^i$, which will be used as the input features in clustering. Although the vectorization of the parameter matrix $\boldsymbol{\Theta}^i$ exhibits an elegant expression of the least squares solution, it may lead to loss of the information contained in the matrix structure of the estimated parameters, when employing the clustering procedure. The $k$-means clustering (or other clustering methods) is a minimization algorithm based on distances between objects. Thus, we propose to convert $\text{vec}(\widehat{\boldsymbol{\Theta}}^i)$ back to a matrix form by writing 
\bqan\label{theta_hat}
\widehat{\boldsymbol{\Theta}}^i=\text{devec}\{({M^i}^TM^i)^{-1}{M^i}^T\mathbf{z}^i\},
\eqan
where ``devec" represents de-vectorization, i.e. arranging $\text{vec}(\widehat{\boldsymbol{\Theta}}^i)$ into an $R\times L$ matrix whose entries correspond to the  parameter matrix $\boldsymbol{\Theta}^i$. This preserves the spatial structure of the columns and rows of the coefficients that correspond to the hills and valleys of the surface. Such spatial structure of the coefficients can be informative when computing the distance between objects, which are based on matrix distance metrics. Next, one aims to find a partition of the set of surfaces $\mathcal{S} = (\mathcal{S}^1, \ldots, \mathcal{S}^n)$ by grouping the set of estimated parameter matrices $\underline{\widehat{\boldsymbol{\Theta}}}^n=\{\widehat{\boldsymbol{\Theta}}^1,\ldots,\widehat{\boldsymbol{\Theta}}^n\}$ so that surfaces in the same cluster have features as similar as possible, and surfaces in different clusters have dissimilar features. That is, for a given number of clusters $K$, the algorithm searches for the set of cluster centers  $\mathbf{c}=\{c_1,\ldots,c_K\}$ that minimizes
\bqan\label{eq.objective}
\frac{1}{n}\sum_{i=1}^n \min_{c\in \mathbf{c}}||\widehat{\boldsymbol{\Theta}}^i-c||,
\eqan
where each $c$ represents an $R\times L$ matrix with real elements, and $||\cdot ||$ is an appropriate matrix norm such as the Frobenius norm.

The $k$-means algorithm finds the partition of the surfaces and their cluster centers $\vc$  in the following iterative manner.

\noindent{\bf Step 1}: Initialize the partitions by setting $\vc^{(0)}$ as $(\widehat{\boldsymbol{\Theta}}^{\ell_1}, \ldots, \widehat{\boldsymbol{\Theta}}^{\ell_K})$ $(\ell_1, \ldots, \ell_K\in \{1, \ldots, n\})$, which can be done, for instance, by choosing\\ 
(a) the $K$ matrices with random entries in the range of  the entries of $\widehat{\boldsymbol{\Theta}}^i\ (i = 1, \ldots, n)$, \\
(b) the $K$ matrices among $\widehat{\boldsymbol{\Theta}}^i\ (i = 1, \ldots, n)$ whose distance (norm) is the largest among themselves: first choose the 2 surfaces that are farthest apart, then sequentially choose other surfaces whose average distance to the previously selected ones is the maximum, \\
(c) the $K$ matrices $(\widehat{\boldsymbol{\Theta}}^{\ell_1}, \ldots, \widehat{\boldsymbol{\Theta}}^{\ell_K})$ among $\widehat{\boldsymbol{\Theta}}^i\ (i = 1, \ldots, n)$ so that the sum of the distances (norm) from each $\widehat{\boldsymbol{\Theta}}^i$ to the closest one in $(\widehat{\boldsymbol{\Theta}}^{\ell_1}, \ldots, \widehat{\boldsymbol{\Theta}}^{\ell_K})$ is  the minimum.\\
(d) $K$ randomly chosen matrices among $\widehat{\boldsymbol{\Theta}}^i\ (i = 1, \ldots, n)$, \\
(e)  the output of a pre-clustering procedure.

\noindent{\bf Step 2}: Assign each surface, i.e. estimated parameter matrix $\widehat{\boldsymbol{\Theta}}^i$, to the closest cluster center $\ell$ according to the minimum distance (norm) $||\widehat{\boldsymbol{\Theta}}^i - \widehat{\boldsymbol{\Theta}}^\ell||\ (\ell \in \{\ell_1, \ldots, \ell_k\})$.

\noindent{\bf Step 3}: Compute the new cluster centers $\vc^{(1)} = (c_1^{(1)}, \ldots, c_K^{(1)})$, where $c_\ell^{(1)}$ is the mean of the matrices $\widehat{\boldsymbol{\Theta}}^i$
for all surfaces $i$ allocated to the $\ell$-th cluster ($\ell=1,\ldots,K$) .

\noindent{\bf Step 4}: Repeat Steps 2 and 3 until there are no more changes in the cluster membership assignments.

\section{Asymptotic Results} \label{sec.theory}

\medskip
\subsection{Strong Consistency without Measurement Errors}

\medskip
In this section we consider the ideal scenario where the entire surface $\mathcal{S}$ is observable without measurement error. Lemma \ref{lemma1} below shows that, given an approximation of $\mathcal{S}$ by a B-spline tensor (projection onto $\mathbb{S}_1\otimes\mathbb{S}_2$), the cluster centers $\mathbf{c}^n$ obtained from minimizing equation (\ref{eq.objective}) converge to a unique (optimal) cluster center set $\mathbf{c}^\ast$, as the number of surfaces $n$ goes to infinity.

Here we use notations similar to those in \cite{Lemaire1983} and \cite{AbrahamEtAl2003}. Let $\Pi(\mathcal{S})$ be the unique matrix $\boldsymbol{\Theta}\in \mathbb{R}^{R\times L}$ such that
\[
\inf_{\boldsymbol{\Theta}\in \mathbb{R}^{R\times L}}||\mathcal{S}-s(\cdot,\boldsymbol{\Theta})|| = ||\mathcal{S}-s(\cdot,\Pi(\mathcal{S}))||.
\]
That is, $\Pi(\mathcal{S})$ is the projection of the smooth surface space $\mathcal{H}^2\otimes\mathcal{H}^2$ onto $\mathbb{S}_1\otimes\mathbb{S}_2$. Let $\mathcal{B}_{\mathbb{R}^{R\times L}}$ and $\mu$ denote the Borel $\sigma$-filed of $\mathbb{R}^{R\times L}$ and the image measure of $P$ induced by $\Pi$. As $\Pi$ is continuous, $(\mathbb{R}^{R\times L}, \mathcal{B}_{\mathbb{R}^{R\times L}},\mu)$ is a probability space. The surface sequence $(\mathcal{S}^1,\ldots,\mathcal{S}^n)$ induces a sequence $\underline{\boldsymbol{\Theta}}^n=(\boldsymbol{\Theta}^1,\ldots,\boldsymbol{\Theta}^n)$ of i.i.d. random matrices $\boldsymbol{\Theta}^i=\Pi(\mathcal{S}^i) \in \mathbb{R}^{R\times L}$.

Let
\begin{eqnarray*}
F&=&\{\mathbf{c}\subset\mathbb{R}^{R\times L}|\text{card}(\mathbf{c})\leq K\},\\
u(\boldsymbol{\Theta},\mathbf{c})&=&\min_{c\in\mathbf{c}}||\boldsymbol{\Theta}-c||_F,
\end{eqnarray*} 
and denote the objective function of the $k$-means algorithm as 
\bqa
u_n(\underline{\boldsymbol{\Theta}}^n,\mathbf{c})&=&\frac{1}{n}\sum_{i=1}^n u(\boldsymbol{\Theta}^i,\mathbf{c}),
\eqa
for all $\boldsymbol{\Theta} \in \mathbb{R}^{R\times L}, c\in\mathbb{R}^{R\times L}$, and $\mathbf{c}\in F$. This differs from the classical clustering methods in that it is composed of norms of matrix differences. Using an appropriate matrix norm, we can establish a result similar to that in \cite{AbrahamEtAl2003}, which we state in Lemma \ref{lemma1}.

%

\begin{lemma} \label{lemma1}
Let $u(\mathbf{c})=\int_{\mathbb{R}^{R\times L}} u(\boldsymbol{\Theta},\mathbf{c})\mu(d\boldsymbol{\Theta})$ and assume that $\text{inf}\{u(\vc)|\vc \in F\} < \text{inf}\{u(\vc)|\vc \in F, card(\vc) < K\}$.
Then, the (unique) minimizer $\mathbf{c}^\ast$ of $u(\cdot)$ exists and there also exists a unique sequence of measurable functions $\mathbf{c}^n$ from $(\Omega, \mathcal{A}, P)$ into $(F, \mathcal{B}_F)$ such that $\mathbf{c}^n(\omega)\subset M_n$ for all $\omega \in \Omega$ and 
\[
u_n(\underline{\boldsymbol{\Theta}}^n, \mathbf{c}^n)=\inf_{\mathbf{c}\subset M_n} u_n(\underline{\boldsymbol{\Theta}}^n, \mathbf{c})\text{ a.s.},
\]
where $\{M_n\}_n$ is an increasing sequence of convex and compact subsets of $\mathbb{R}^{R\times L}$ such that $\mathbb{R}^{R\times L}=\cup_n M_n$.
Furthermore, this sequence $\{\mathbf{c}^n\}$ is strongly consistent to $\mathbf{c}^\ast$ with respect to the Hausdorff metric.
\end{lemma}

\subsection{Strong Consistency with Measurement Errors}

Consider the more realistic model specified in equation \eqref{model}, where the surfaces are actually recorded with some measurement errors. Given a set of observations, $\{(x_1^i, y_1^i, z_1^i), \ldots, (x_{m_i}^i, y_{m_i}^i, z_{m_i}^i)\}$, one can estimate the surface $s^i(\cdot,\boldsymbol{\Theta})$ by the B-spline estimate $\hat{s}^i = s(\cdot,\widehat{\boldsymbol{\Theta}})$, where $\widehat{\boldsymbol{\Theta}}$ is least-square estimated B-spline coefficient matrix given in equation \eqref{theta_hat}. Assume that each surface is observed at different grid points over a compact set $[a,b]\times[c,d]$, for some real constants $a, b, c,$ and $d$ $(a<b, c<d)$. Assume also that for each $i$, $x_1^i, \ldots, x_{m_i}^i$ and $y_1^i, \ldots, y_{m_i}^i$ are i.i.d. with probability distributions $h$ and $g$ respectively. 
The following lemma shows that the estimator $\widehat{\boldsymbol{\Theta}}$ is strongly consistent for $\boldsymbol{\Theta}=\Pi(\mathcal{S})$, projection of the smooth surface space $\mathcal{H}^2\otimes\mathcal{H}^2$ onto the space generated by the B-spline bases.

\begin{lemma}\label{lemma2}
Assume the spline bases functions $B_{x,1}, \ldots, B_{x,R}$ and $B_{y,1}, \ldots, B_{y,L}$ are linearly independent on the support of $h$ and $g$ respectively. Assume also that the surfaces $\mathcal{S}^i$ belong to the space $\mathbf{S}$ defined as $\mathcal{H}^2\otimes\mathcal{H}^2$ restricted to bounded variation on $[a,b]\times[c,d]$.
Then, $\widehat{\boldsymbol{\Theta}}$ converges strongly to $\boldsymbol{\Theta}=\Pi(\mathcal{S})$ when $m\rightarrow\infty$ uniformly over space $\mathbf{S}$. As a result, for almost all $\omega\in\Omega$ and all $\mathcal{S}\in\mathbf{S}$, $||\widehat{\boldsymbol{\Theta}}-\boldsymbol{\Theta}||\rightarrow 0$ as $m$ goes to infinity.
\end{lemma}

The proofs of Lemma \ref{lemma1} and Lemma \ref{lemma2} are similar to those in \cite{AbrahamEtAl2003} in curve clustering. For the details of the proofs, please refer to the appendix.

\section{Generalization to Tensor Clustering} \label{sec.tensor}

The framework designed in Section \ref{sec.method} can be generalized to multi-dimensional clustering. Denote $\mathcal{S}^i(\vx), \vx = (x_1, \ldots, x_d) \in \mathcal{Q}$, as the data generating tensor mechanism, where $\mathcal{Q}$ is a subset of $\mathbb{R}^d$ and $d$ is the tensor dimension $(d\geq 2)$. The observed data $(z_i, \vx_i)$ is such that
\bqan\label{modelGeneral}
z_j^i = \mathcal{S}^i(\vx_j^i)+\epsilon_j^i,\quad \ (1\leq j\leq m_i;1\leq i\leq n).
\eqan

The approximation of surface $\mathcal{S}^i(\vx)$ is then based on the smooth B-Spline $d$-dimensional tensor product
\bqa
s^i(\vx,\boldsymbol{\Theta})=\sum_{r_1=1}^{R_1}\ldots\sum_{r_d=1}^{R_d} B_{x_1,r_1}(x_1)\ldots B_{x_d,r_d}(x_d)\theta_{r_1, \ldots, r_d}^i,
\eqa
where $\theta_{r_1, \ldots, r_d}^i$ are coefficients to be estimated and $B_{x_1,r_1}(\cdot), \ldots, B_{x_d,r_d}(\cdot)$ are B-spline basis functions that generate the spline spaces $\mathbb{S}_1 = \text{span}\{B_{x_1,1}, \ldots, B_{x_1,R_1}\}$, $\ldots$, $\mathbb{S}_d = \text{span}\{B_{x_d,1}, \ldots, B_{x_d,R_d}\}$ respectively. The array of coefficients $\boldsymbol{\Theta}$ has dimension $\prod_{i=1}^d r_i$, which is the number of parameters to be estimated.

The multi-dimensional  space of smooth surfaces in $\mathcal{H}^2\otimes\ldots\otimes\mathcal{H}^2$ is then approximated by $\mathbb{S}_1\otimes\ldots\otimes\mathbb{S}_d$.
The least squares solution of this model can be written as
\begin{eqnarray*}
\text{vec}(\widehat{\boldsymbol{\Theta}}^i) &=& \arg\min_{\boldsymbol{\Theta}}\sum_{j=1}^{m_i}[z_j^i-s^i(\vx_{j}^i,\boldsymbol{\Theta})]^2\\
&=&\arg\min_{\boldsymbol{\Theta}}\sum_{j=1}^{m_i}[z_j^i-({\boldsymbol{B}^i_{x_d}(x_{dj}^i)}\otimes\ldots\otimes{\boldsymbol{B}^i_{x_1}(x_{1j}^i)})^T\text{vec}(\boldsymbol{\Theta})]^2\\
&=&({M^i}^TM^i)^{-1}{M^i}^T\mathbf{z}^i,
\end{eqnarray*}
where $\text{vec}(\boldsymbol{\Theta}^i)$ is the vectorization of matrix $\boldsymbol{\Theta}^i$ arranged by columns,   $\mathbf{z}^i=(z_1^i,\ldots,z_{m_i}^i)^T$, ${\boldsymbol{B}^i_x}(x_j^i)=(B_1^i(x_j^i),\ldots,B_R^i(x_j^i))^T$, ${\boldsymbol{B}^i_y}(y_j^i)=(B_1^i(y_j^i),\ldots,B_L^i(y_j^i))^T$, and $M^i$ is the $m_i\times RL$ matrix with its $j$th row equal to the $1\times RL$ vector of $(\boldsymbol{B}^i_y(y_j^i)\otimes \boldsymbol{B}^i_x(x_j^i))^T$. The proposed procedure can be applied to $\widehat{\boldsymbol{\Theta}}^i$ in a similar fashion as in Section \ref{sec.method}, except that one needs to employ an appropriate array norm to evaluate the distances between $\widehat{\boldsymbol{\Theta}}^i$ and $\widehat{\boldsymbol{\Theta}}^j$ $(i\ne j; i,j=1,...,n)$. We omit the details for the discussion of tensor clustering in this paper.

\section{Simulation Study}\label{sec.simul}

\indent \indent In this section we investigate the finite sample performance of the proposed method in clustering surfaces through several simulation scenarios. For comparison purposes, we also evaluate the performance of the $k$-means clustering, the benchmark procedure, which does not allow the use of random grid coordinates. 
For both the proposed method and the benchmark $k$-means we chose the initial guess of the cluster centers in the first step of our proposed algorithm as follows: consider the possible initial guesses given by initialization methods (b), (c), and 50 random initializations as described in method (d) of the $k$-means algorithm in Section \ref{sec.method}. From these 53 possible initial guesses, we choose the one whose $K$ matrices of estimated coefficients, when defined as the center of clusters, have the minimum average distance to the objects in the data assigned to their corresponding clusters. In our numerical studies, we focus on the Frobenius norm as an example

The first simulation setting concerns two clusters whose cluster centers are the following (probability density) surfaces, each composed of a mixture of Normal distributions:
{\small
\bqa
&& f_1(x,y) = 0.3\phi\left((x,y);\begin{pmatrix}
0 \\
-3
\end{pmatrix},\begin{pmatrix}
1 & 0\\
0 & 5
\end{pmatrix}\right) + 0.7\phi\left((x,y);\begin{pmatrix}
0 \\
3
\end{pmatrix}, \begin{pmatrix}
1 & 0\\
0 & 1
\end{pmatrix}\right)\\
&& f_2(x,y) = 0.3\phi\left((x,y); \begin{pmatrix}
0 \\
-3
\end{pmatrix}, c\begin{pmatrix}
1 & 0\\
0 & 1
\end{pmatrix}\right) + 0.7\phi\left((x,y);\begin{pmatrix}
0 \\
3
\end{pmatrix},  c\begin{pmatrix}
1 & 0\\
0 & 1
\end{pmatrix}\right)
\eqa
}where $\phi((x,y);\mu, \Sigma)$ is the probability density function of a bivariate Normal distribution with mean $\mu$ and variance $\Sigma$, and $c$ is a fixed constant. Note that the covariance matrix of each Normal component of the second cluster center is a multiple of $c$. For small values of $c$, the peaks, or modes, of the Normal mixture densities are very high. As $c$ increases, the peaks become more and more flat.  The simulations we present below show the performance of the proposed clustering procedure when the constant $c$ varies, that is, with varying degrees of difficulty. Figure \ref{fig.Scenario1} shows the centroid surfaces from both clusters, where the top plots display the surface of the second cluster for $c = 0.2, 1$ and 3. Note that distinguishing between these two clusters when $c = 1$ is challenging, due to the fact that the only difference is the variance of the first component of the normal mixture. The task of clustering becomes even harder with the presence of random error in the data generating process which we describe next.

\begin{figure} 
  \centering
      \includegraphics[width=0.9\textwidth]{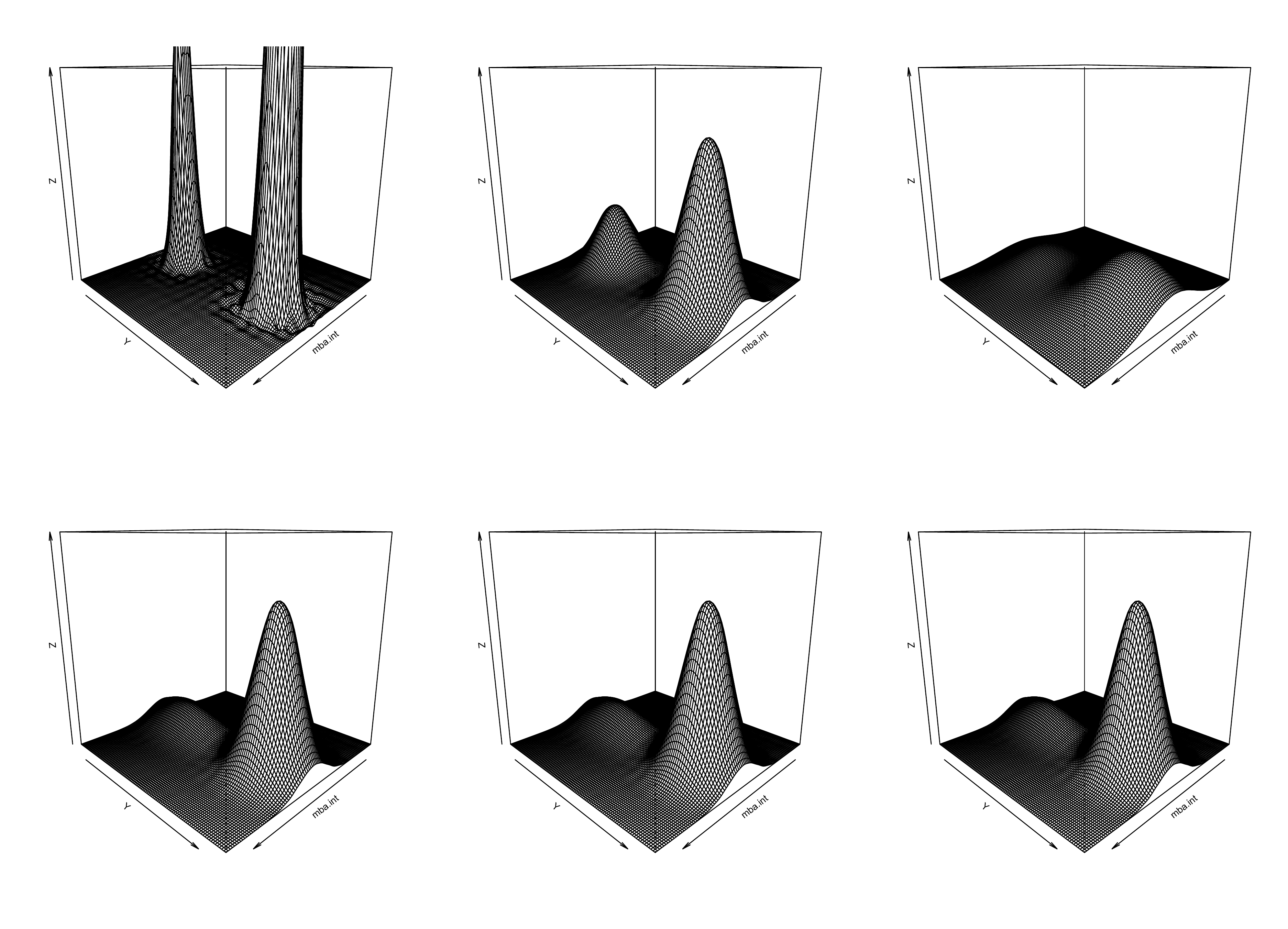}
  \caption{Top: from left to right, surface of the second cluster center (density $f_2$) with constant $c$ equal to 0.2, 1, and 3 respectively. Bottom: surface of the first cluster center (density $f_1$) for ease of visual comparison with the second cluster center.}\label{fig.Scenario1}
\end{figure}

The simulated data $(x, y, z)$ for each surface were generated in a 20 $\times$ 20 grid, i.e. a total of 400 data points, in the square region $(-5,5)\times(-5,5)$. The data for Clusters 1 and 2 are generated as follows
{\small
\bqa
&&\text{Cluster 1}: \{(x, y, z): x_j = y_j = -5 + j/2, j = 1, \ldots, 20; z_j = f_1(x_j,y_j) + \epsilon_j^1 \}\\
&&\text{Cluster 2}: \{(x, y, z): x_j = y_j = -5 + j/2, j = 1, \ldots, 20; z_j = f_2(x_j,y_j) + \epsilon_j^2 \},
\eqa}where $\epsilon_j^1$ and $\epsilon_j^2$ are i.i.d. random errors from $N(0, 0.015^2)$ and $N(0, 0.01^2)$ respectively. We generated a total of $n = 60$ surfaces, with 30 samples belonging to each of the two clusters.  In the simulations in this paper, we used cubic B-spline basis and a fixed number of 6 knots in each axis for the estimation of the surfaces. This process is repeated for $B = 500$ Monte Carlos simulation runs. 

It is well known that the initialization of the $k$-means procedure can have an immense influence on the clustering results (\cite{PenaEtAl1999}, \cite{FrantiSieranoja2019}). For this reason and for a fair comparison between the proposal and the benchmark, we initialized each procedure in the same way as described in the first paragraph of  this section. The benchmark applies the $k$-means algorithm to the vectorized  raw data set, while the proposal employs the $k$-means to the B-spline estimated coefficients. 

In order to evaluate the results we consider the following performance measure. Let $\mathcal{S}^{i(b)}$ denote the randomly generated surface $i\ (i = 1, \ldots, n)$ in the $b$-th simulation run $(b = 1, \ldots, B)$.
Let $L(\mathcal{S}^{i(b)}), L^*(\mathcal{S}^{i(b)}) \in \{1, \ldots, K\}$ be the predicted and true cluster membership for surface $i$ respectively.
We compute
\bqa
\label{errorrate}
\phi &=& \frac{\sum_{b=1}^{B}\underset{\tau \in T}{\min}\sum_{i=1}^{n}I(L(\mathcal{S}^{i(b)}) \neq \tau(L^*(\mathcal{S}^{i(b)})))}{B\nu},\text{ where}\\
\nu &=& \frac{\sum_{b=1}^{B}I\left(\underset{\tau \in T}{\min}\sum_{i=1}^{n}I(L(\mathcal{S}^{i(b)}) \neq \tau(L^*(\mathcal{S}^{i(b)})))\geq 1\right)}{B},\\
\eqa
and the $T$ in ``$\tau\in T$" is the set of permutations over $\{1, \ldots, K\}$. Note that the term $\min_{\tau \in T}\sum_{i=1}^{n}I(L(\mathcal{S}^{i(b)}) \neq \tau(L^*(\mathcal{S}^{i(b)})))$ is the number of mis-specification errors of a clustering procedure, which is based on the MCE measure in Fraiman et al. (2013).
Hence, the quantity $\nu$ measures the proportion of times that the algorithm mis-specifies, that is, the proportion of simulation runs with at least one surface being assigned to an incorrect cluster. Hence, the performance measure $\phi$ is the mean mis-specification, i.e., the average number of surfaces being assigned to incorrect clusters in the simulations with at least one mis-specification.

Figure \ref{fig.comparison1} shows the results of the proposed clustering method and the benchmark in this first scenario. For small values of $c$ (from 0.2 to 0.5) the variance of the densities in the first cluster is small and hence the hill is high (see top and bottom left plots of Figure \ref{fig.Scenario1}). In this case both algorithms cluster the data without any incorrectly grouped surfaces. While at $c = 0.7$ the benchmark $k$-means incorrectly mis-specifies an average of about 18 curves ($\phi = 18.8$), the proposed method can still keep this rate low at $\phi = 5.3$. For values of $c$ from 1 to 1.7, the two bivariate densities that compose each cluster are very similar (see top and bottom middle plots in Figure \ref{fig.Scenario1}), and both algorithms have similar performance by incorrectly clustering about 20 to 25 surfaces out of 60. For larger values of $c$ (greater than 2), the variance of the densities in cluster 1 is large, so that the hills are low and the difference between the clusters are again more visible (see top and bottom right plots in Figure \ref{fig.Scenario1}). For these values of $c$ the proposed procedure again yields very low number of incorrectly clustered surfaces, while the benchmark $k$-means still struggles to correctly cluster surfaces until $c$ is very large.

\begin{figure}
  \centering
      \includegraphics[width=0.5\textwidth]{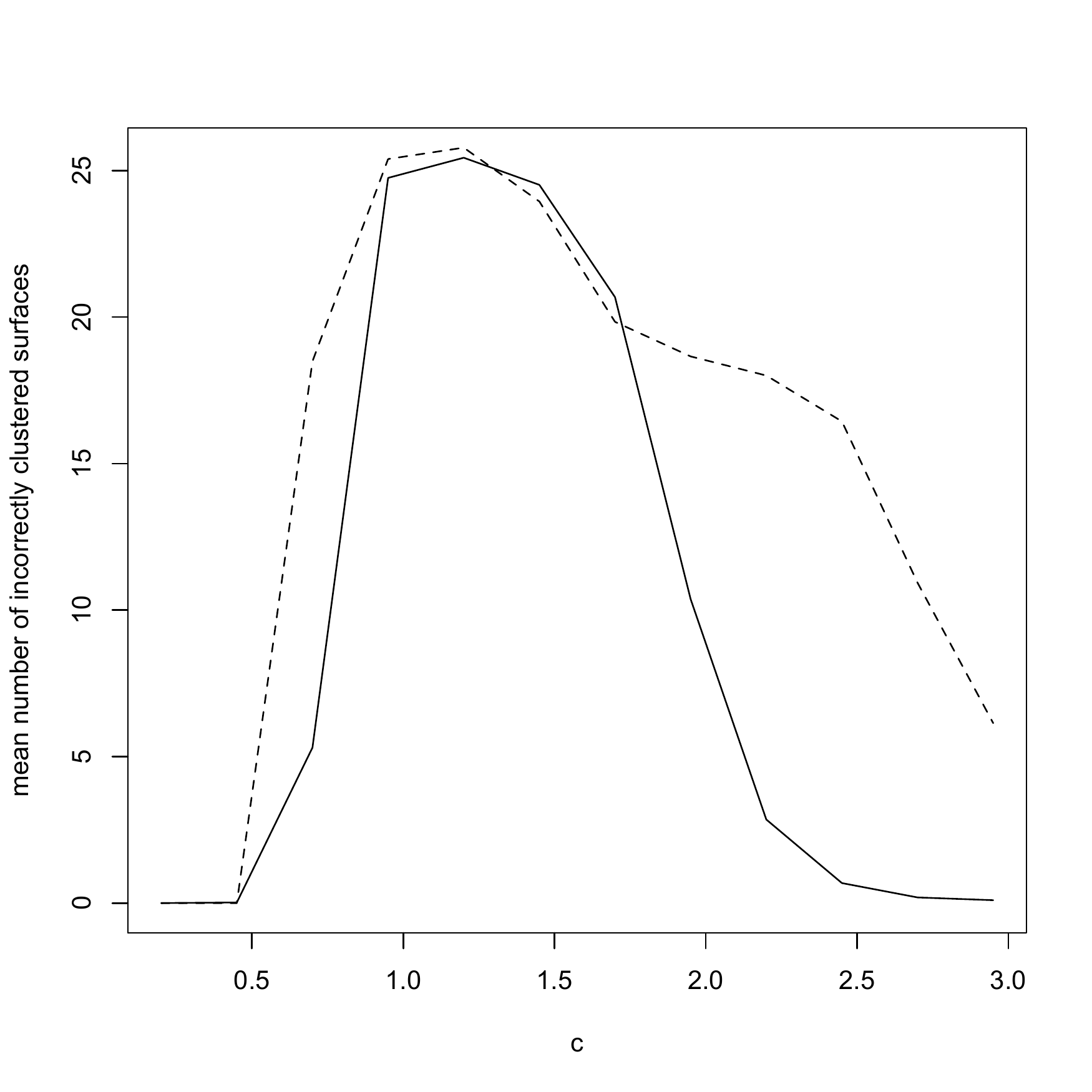}
  \caption{Comparison of mean number of incorrectly clustered surfaces, out of $n=60$,  with varying values of $c$ in the first simulation setting. Dashed curve represents the benchmark, and the solid curve corresponds to the proposed method.}\label{fig.comparison1}
\end{figure}

The second simulation setting is composed of 3 clusters whose centroid surfaces are defined by
\bqa
&& f_3(x,y) = 0.3\phi\left((x,y);\begin{pmatrix}
0 \\
-3
\end{pmatrix}, \begin{pmatrix}
1 & 0\\
0 & 1
\end{pmatrix}\right) + 0.7\phi\left((x,y);\begin{pmatrix}
0 \\
1
\end{pmatrix}, \begin{pmatrix}
1 & 0\\
0 & 1
\end{pmatrix}\right)\\
&& f_4(x,y) = 0.3\phi\left((x,y);\begin{pmatrix}
0 \\
1
\end{pmatrix},  c\begin{pmatrix}
1 & 0\\
0 & 2
\end{pmatrix}\right) + 0.7\phi\left((x,y); \begin{pmatrix}
0 \\
-3
\end{pmatrix},  c\begin{pmatrix}
1 & 0\\
0 & 1
\end{pmatrix}\right)\\
&& f_5(x,y) = 0.3\phi\left((x,y);\begin{pmatrix}
0 \\
-2
\end{pmatrix},  c\begin{pmatrix}
1 & 0\\
0 & 1
\end{pmatrix}\right) + 0.7\phi\left((x,y);\begin{pmatrix}
1 \\
0
\end{pmatrix},  c\begin{pmatrix}
1 & 0\\
0 & 1
\end{pmatrix}\right).
\eqa
Figure \ref{fig.Scenario3} shows the centroid surfaces from the 3 clusters for $c = 0.2, 1$ and $3$. The first two clusters are easier to distinguish since the hills are opposites. However, the third cluster may bring a challenge for values of $c$ near 1, especially when there is random error.

\begin{figure}[http!]
  \centering
      \includegraphics[width=0.9\textwidth]{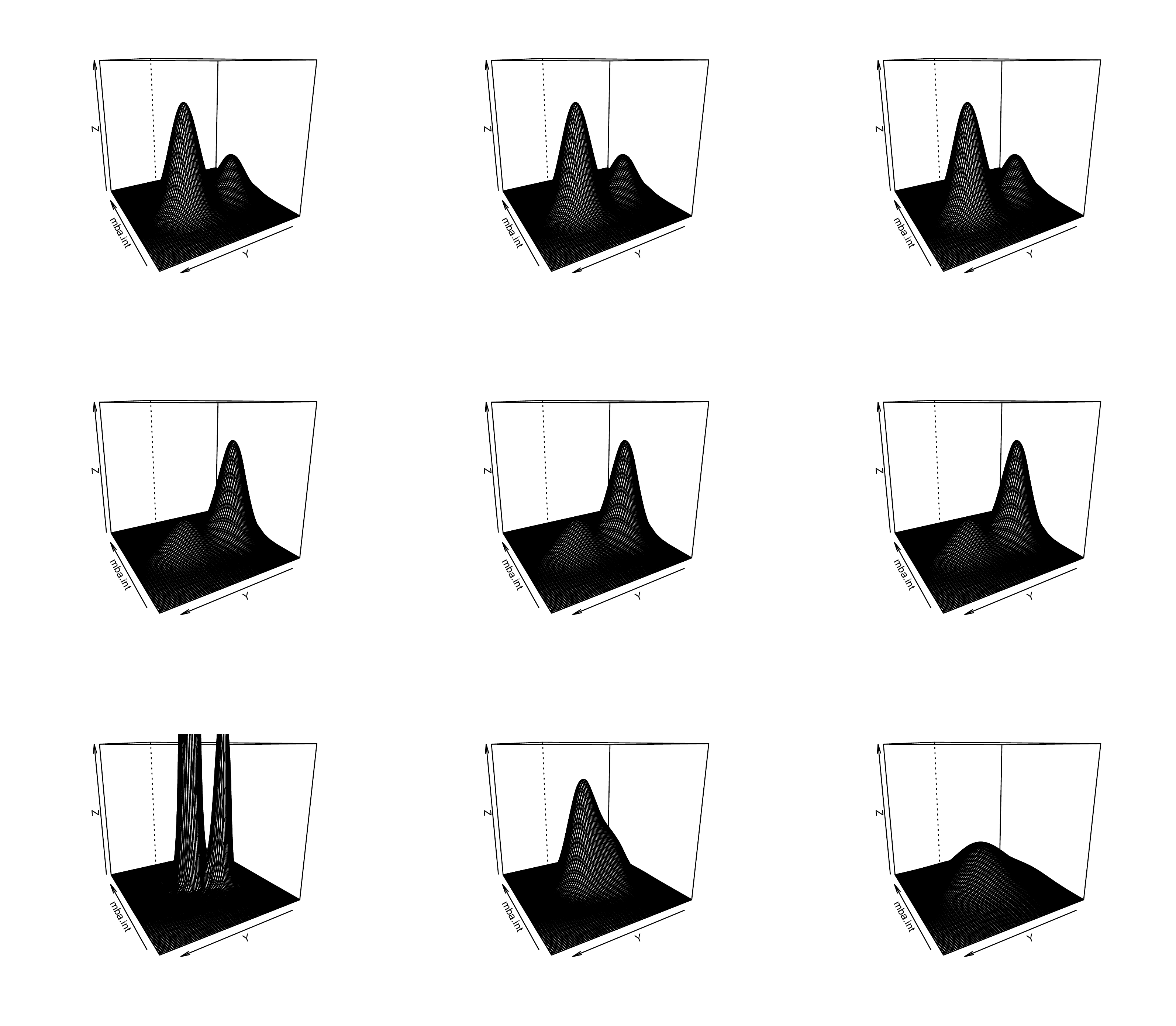}
  \caption{Top: from left to right, surface of the second cluster center with constant $c$ equal to 0.2, 1, and 3 respectively. Bottom: surface of the first cluster center for ease of visual comparison with the second cluster center.}\label{fig.Scenario3}
\end{figure}

The data for each cluster are simulated as follows:
\bqa
&&\text{Cluster 1}: \{(x, y, z): x_j = y_j = -5 + j/2, j = 1, \ldots, 20; z_j = f_3(x_j,y_j) + \epsilon_j^1 \}\\
&&\text{Cluster 2}: \{(x, y, z): x_j = y_j = -5 + j/2, j = 1, \ldots, 20; z_j = f_4(x_j,y_j) + \epsilon_j^2 \},\\
&&\text{Cluster 3}: \{(x, y, z): x_j = y_j = -5 + j/2, j = 1, \ldots, 20; z_j = f_5(x_j,y_j) + \epsilon_j^3 \},
\eqa
where $\epsilon_j^1, \epsilon_j^2, \epsilon_j^3$ are i.i.d. random errors from $N(0, 0.015^2)$. We generated a total of $n = 60$ surfaces, where 20 surfaces belong to each cluster. This process was repeated for $B = 500$ Monte Carlos simulation runs. Figure \ref{fig.comparison2} displays the comparison in terms of the average number of incorrectly clustered surfaces for different values of $c$. The results suggest that the proposed method consistently outperforms the benchmark for $0<c<3.0$, especially for small values of $c$. For large values of $c$ the proposed algorithm has an increase in mis-specification error, possibly due to the random nature of the data generated from the small hill in the third cluster that lies between the hills in the densities of clusters 1 and 2. The randomness of the data from the third cluster may sometimes lead to estimated surfaces with slightly higher hills in positions where the other clusters have hills, bringing a challenge to the clustering methods.

\begin{figure}[http!]
  \centering
      \includegraphics[width=0.5\textwidth]{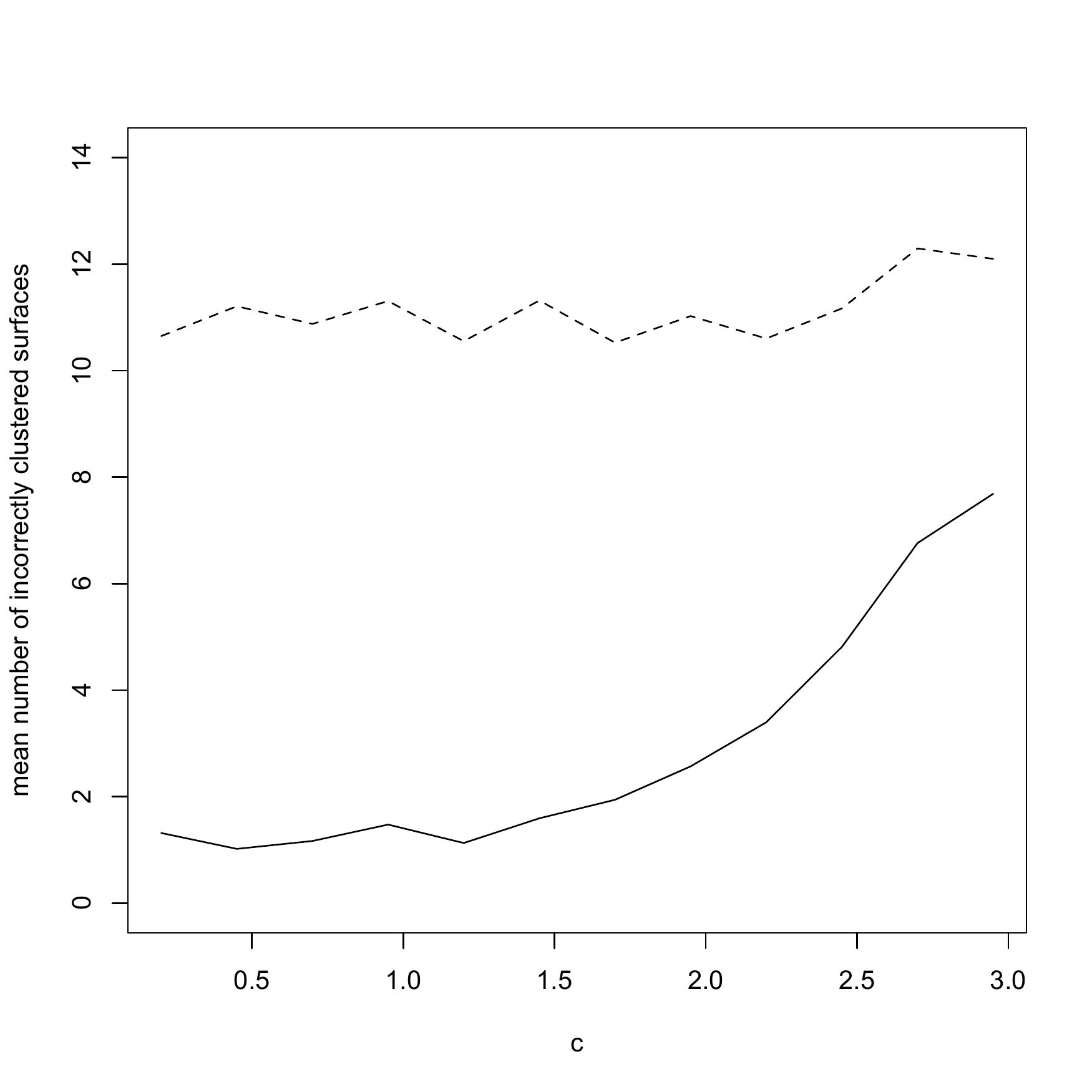}
  \caption{Comparison of mean number of incorrectly clustered surfaces, out of $n=60$, with varying values of $c$ in the second simulation setting. Dashed curve represents the benchmark, and the solid curve corresponds to the proposed method.}\label{fig.comparison2}
\end{figure}

\section{Real Data Analysis: EEG Clustering}\label{sec.data.analysis}

In this section we illustrate the proposed surface clustering method in an analysis of the Electroencephalogram (EEG) dataset, which is available at the University of California Irvine Machine Learning Repository\\ (https://archive.ics.uci.edu/ml/datasets/eeg+database). The dataset is composed of 122 subjects that are divided into two groups: alcohol and control. 
Each subject was exposed to either a single stimulus (S1) or  two stimuli (S1 and S2). The stimulus was a picture of an object chosen from the 1980 Snodgrass and Vanderwart picture set. For the case of two stimuli, they were either matched, where S1 was identical to S2, or not matched, where S1 was different from S2.
On the scalp of each subject 64 electrodes were positioned according to the Standard Electrode Position Nomenclature, American Electroencephalographic Association, and measurements were taken at 256 Hz (3.9-msec epoch) for 1 second (See \cite{ZhangEtAl1995} for details).
The channels (i.e. electrodes) and time compose the surface domain $(x, y)$ and the measurements are the response $z$. 

We averaged over all trials for each individual and used the mean surface to represent the output of each subject's response to the stimuli. 
Visualization of the raw data  of a subject in the control group as well as the raw data of a subject in the alcohol group are shown in Figure \ref{fig.EEG}.  
\begin{figure}[http!]
  \centering
      \includegraphics[width=1.1\textwidth]{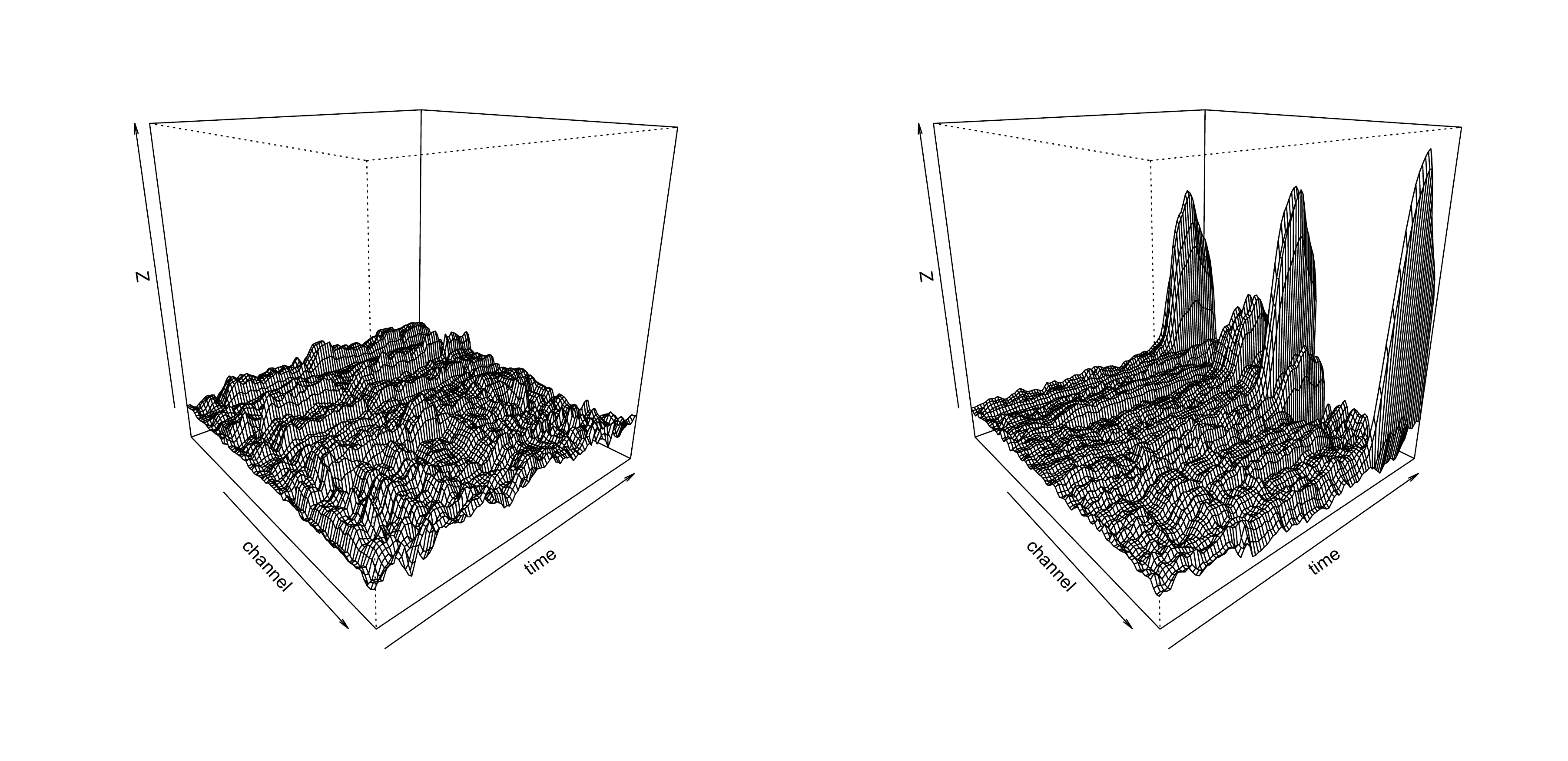}
  \caption{Example of EEG measurements for a subject in the control group (left) and alcohol group (right). }\label{fig.EEG}
\end{figure}
The EEG surface of the control subject seems to be somewhat flat with some random noise, while the EEG surface of the subject in the alcohol group shows a few hills for some channels at later time stamps. However, this is not the case for all subjects. It can also be observed that the EEG surfaces of some subjects in the control group have hills at later time stamps, while some subjects in the alcohol group  have somewhat flat EEG surfaces throughout time. This imposes practical challenges on the clustering task, as we discuss below.

We applied our proposed clustering method to the full dataset of 122 patients, where the elements clustered by the $k$-means algorithm were the mean surfaces of the subjects across their trials. Given that we know the actual groups of the subjects, i.e. either control or alcohol, we can compute how many mistakes were made by the proposed algorithm when applied to this dataset. The number of incorrectly clustered subjects was 51 (41.8\%) when using surfaces from S1,  41 (33.6\%) for S2 match, and 38 (31.1\%) for S2  no match. Although an accuracy rate of about 60 to 70\% correctly clustered subjects is not very high, we recall the fact that having hills at later time stamps is not a feature that only belongs to EEG surfaces of subjects in one particular group.

\section{Conclusions}

This article proposes a new approach for clustering surfaces using the coefficients obtained from B-splines that approximate their data generating smooth forms as the basis for a $k$-means clustering algorithm. The proposed method is shown to be strongly consistent in both stochastic and non-stochastic cases. Compared to the classical $k$-means procedure applied directly to the vectorized version of the data that may lose the geometric structure of the surface, our surface clustering  procedure performs consistently better in correctly clustering surfaces observed with noise in simulation studies. From the wide range of applications of surface clustering, in this paper we studied the identification of effects of alcohol in the brain by clustering Electroencephalogram data from 122 patients and clustering into alcohol and control groups, where the surfaces were defined by the stimuli response at 64 electrodes throughout  time stamps. 

\section{Acknowledgements}

We would like to thank Fapesp for partially funding this research (Fapesp 2018/04654-9, 2017/15306-9, 2019/04535-2).

\section{Appendix}

{\bf Proof of Lemma \ref{lemma1}}
\begin{proof}
Following similar steps as in the proof of Proposition 1 in Abraham et al. (2003), we need to show $u(\boldsymbol{\Theta}, \cdot)$ is strictly convex so that $u_n(\underline{\boldsymbol{\Theta}}^n,\cdot)$ and $u(\cdot)$ have unique minimizers if they exist.

Define the function $\phi$ from $F$ into $(\mathbb{R}^{R\times L})^K$ such that $\phi(\mathbf{c})$ is an ordered vector of elements $c_i$ in $\mathbf{c}$. Suppose we denote the order using the notation $\leq$ where $c_i\leq c_j$ means $c_j$ is ordered in front of $c_i$. Let function $\psi$ be one from $(\mathbb{R}^{R\times L})^K$ to $F$ such that $\psi(c_1,\ldots,c_K)=\{c_1,\ldots,c_K\}$. Thus, $\psi(\phi(\mathbf{c}))=\mathbf{c}$. 

Let $\lambda$ be a scalar. For two elements $\mathbf{c}$ and $\mathbf{c}^\prime$ from $F$, define scalar multiplication and sum of elements as
\begin{eqnarray*}
\lambda\mathbf{c}&:=&\psi(\lambda\phi(\mathbf{c}))=\{\lambda c_1,\ldots,\lambda c_k\},\\
\mathbf{c}+\mathbf{c}^\prime&:=& \psi(\phi(\mathbf{c})+\phi(\mathbf{c}^\prime)),
\end{eqnarray*}
so that $\lambda\mathbf{c} \in F$ and $\mathbf{c}+\mathbf{c}^\prime \in F$.
Then
\begin{eqnarray*}
u(\boldsymbol{\Theta},\lambda \mathbf{c}+(1-\lambda)\mathbf{c}^\prime)&=&\min_{c\in\lambda \mathbf{c}+(1-\lambda)\mathbf{c}^\prime}||\boldsymbol{\Theta}-c||\\
&=&\min_{1\leq i\leq K}||\boldsymbol{\Theta}-(\lambda c_i+(1-\lambda)c_i^\prime||\\
&<&\min_{1\leq i\leq K}\lambda||\boldsymbol{\Theta}-c_i||+(1-\lambda)||\boldsymbol{\Theta}-c_i^\prime||\\
&\leq & \lambda u(\boldsymbol{\Theta},\mathbf{c})+(1-\lambda)u(\boldsymbol{\Theta},\mathbf{c}^\prime).
\end{eqnarray*}
The inequality in the third line only uses the property of triangle inequality of a matrix norm, so that any appropriate matrix norm can be used.

Therefore, $u(\boldsymbol{\Theta}, \cdot)$ is strictly convex, which implies that $u_n(\underline{\boldsymbol{\Theta}}^n,\cdot)$ and $u$ are also strictly convex. Consequently, the minimizers of $u_n(\underline{\boldsymbol{\Theta}}^n,\cdot)$ and $u$ are unique if they exist. The existence of the minimizer and the strong consistency of $\mathbf{c}^n$ to $\mathbf{c}^\ast$ follow from proposition 7, and theorems 1 and 2 in \cite{Lemaire1983}.
\end{proof}

{\bf Proof of Lemma \ref{lemma2}}
\begin{proof}
For ease of notation, we omit the superscript $i$ in the proof of this lemma. Here we follow steps similar to those in the proof of Proposition 1 in \cite{AbrahamEtAl2003}.

 Let $P_\epsilon$ denote the distribution of the error, and recall that $h$ is the bivariate distribution of $(x_j, y_j)$. Following the notations and set-up in Appendix 2 of Abraham et al. (2003), denote $z(x,y,\epsilon)=g(x,y)+\epsilon$. Thus,
\begin{eqnarray*}
\|g(x,y)\|^2&=&\int g^2(x,y)h(d(x,y)),\\
\|z-s(\cdot,\boldsymbol{\Theta})\|^2&=&\int (g(x,y)+\epsilon-s(x,y,\boldsymbol{\Theta}))^2h(d(x,y))P_\epsilon(d\epsilon)=\|g-s(\cdot,\boldsymbol{\Theta})\|^2+\|\epsilon\|^2\\
\|g\|_m^2&=&\frac{1}{m}\sum_{j=1}^m g^2(x_j,y_j)\\
\|z-s(\cdot,\boldsymbol{\Theta})\|_m^2&=&\frac{1}{m}\sum_{j=1}^m[g(x_j,y_j)+\epsilon_j-s(x_j,y_j,\boldsymbol{\Theta})]^2=\|\epsilon-(g-s(\cdot,\boldsymbol{\Theta}))\|_m^2.
\end{eqnarray*}

Using similar arguments, we can obatain uniform strong law of large numbers on the space $\mathcal{F}=\{z-s(\cdot,\alpha),g\in\mathbf{S}, s\in \mathbf{B}\}$, where $\mathbf{S}$ is the space for all the surfaces and $\mathbf{B}$ is the matrix subspace generated by the B-spline basis. That is,
\begin{equation}
\label{eq:lln}
\sup_{g\in\mathbf{S},s(\cdot,\alpha)\in\mathbf{B}}|\|z-s(\cdot,\alpha)||_m^2-||z-s(\cdot,\alpha)\|^2|\rightarrow 0\text{ almost surely}.
\end{equation}

Fix $\eta>0$. By equation \eqref{eq:lln} for almost every $\omega\in\Omega$ and all $g\in\mathbf{S}$, there exists an integer $N$ such that for any $m>N$, $\sup_{s(,\boldsymbol{\Theta})}|\|z-s(\cdot,\boldsymbol{\Theta})\|_m^2-\|z-s(\cdot,\boldsymbol{\Theta})^2|<\eta/2$. Let $\widehat{\boldsymbol{\Theta}}$ be the least-square estimate of $\Pi(g)$ using observations $z_j$ at the design points $(x_j,y_j)$. Then, for sufficiently large $m$,
\begin{eqnarray*}
\|\epsilon\|^2+\|g-s(\cdot,\widehat{\boldsymbol{\Theta}})\|^2&=&\|z-s(\cdot,\widehat{\boldsymbol{\Theta}})\|^2\\
&\leq & \|z-s(\cdot,\widehat{\boldsymbol{\Theta}})\|_m^2+\frac{\eta}{2}\\
&\leq & \|z-s(\cdot,\Pi(\mathcal{S}))\|_m^2+\frac{\eta}{2}\\
&\leq & \|g+\epsilon-s(\cdot,\Pi(\mathcal{S}))\|^2+\eta\\
&=&\|\epsilon\|^2+\|g-s(\cdot,\Pi(\mathcal{S}))\|^2+\eta,
\end{eqnarray*}
which implies
\[
\|s(\cdot,\Pi(\mathcal{S}))-s(\cdot,\widehat{\boldsymbol{\Theta}})\|^2\leq \eta.
\]
This concludes the proof.

\end{proof}

\bibliographystyle{apalike} 
\bibliography{mybibfile}

\begin{thebibliography}{}

\bibitem[Abraham et~al., 2003]{AbrahamEtAl2003}
Abraham, C., Cornillon, P.~A., Matzner-Løber, E., and Molinari, N. (2003).
\newblock Unsupervised curve clustering using b-splines.
\newblock {\em Scandinavian Journal of Statistics}, 30(3):581--595.

\bibitem[Abu-Jamous et~al., 2015]{Abu-Jamous15}
Abu-Jamous, B., Fa, R., and Nandi, A. (2015).
\newblock {\em Interactive cluster analysis in bioinformatics}.
\newblock Wiley.

\bibitem[Abualigah et~al., 2018]{Abualigah18}
Abualigah, L.~M., Khader, A.~T., and Hanandeh, E.~S. (2018).
\newblock A combination of objective functions and hybrid krill herd algorithm
  for text document clustering analysis.
\newblock {\em Engineering Applications of Artificial Intelligence},
  73:111--125.

\bibitem[Blackhurst et~al., 2018]{Blackhurst18}
Blackhurst, J., Rungtusanatham, M.~J., Scheibe, K., and Ambulkar, S. (2018).
\newblock Supply chain vulnerability assessment: A network based visualization
  and clustering analysis approach.
\newblock {\em Journal of Purchasing and Supply Management}, 24:21--30.

\bibitem[Boullé, 2012]{BoulleEtAl2014}
Boullé, M. (2012).
\newblock Functional data clustering via piecewise constant nonparametric
  density estimation.
\newblock {\em Pattern Recognition}, 45(12):4389 -- 4401.

\bibitem[{de Boor}, 1971]{deBoor1971}
{de Boor}, C. (1971).
\newblock Subroutine package for calculating with b-splines.
\newblock {\em Techn.Rep. LA-4728-MS, Los Alamos Sci.Lab, Los Alamos NM}, pages
  109 -- 121.

\bibitem[{de Boor}, 1972]{deBoor1972}
{de Boor}, C. (1972).
\newblock On calculating with b-splines.
\newblock {\em Journal of Approximation Theory}, 6(1):50 -- 62.

\bibitem[de~Boor, 1977]{Boor1977}
de~Boor, C. (1977).
\newblock Package for calculating with b-splines.
\newblock {\em SIAM Journal on Numerical Analysis}, 14(3):441--472.

\bibitem[Dias and Gamerman, 2002]{DiasGamerman2002}
Dias, R. and Gamerman, D. (2002).
\newblock A {B}ayesian approach to hybrid splines nonparametric regression.
\newblock {\em Journal of Statistical Computation and Simulation.},
  72(4):285--297.

\bibitem[Duran et~al., 2019]{Duran19}
Duran, A.~H., Greco, T.~M., Vollmer, B., M., C.~I., Crunewald, K., and Topf, M.
  (2019).
\newblock Protein interactions and consensus clustering analysis uncover
  insights into herpesvirus virion structure and function relationships.
\newblock {\em {PLOS} Biology}.

\bibitem[Febrero-Bande and de~la Fuente, 2012]{BandeFuente2012}
Febrero-Bande, M. and de~la Fuente, M. (2012).
\newblock Statistical computing in functional data analysis: The r package
  fda.usc.
\newblock {\em Journal of Statistical Software, Articles}, 51(4):1--28.

\bibitem[Ferraty and Vieu, 2006]{FerratyVieu2006}
Ferraty, F. and Vieu, P. (2006).
\newblock {\em Nonparametric functional data analysis}.
\newblock Springer Series in Statistics.

\bibitem[Floriello, 2011]{Floriello2011}
Floriello, D. (2011).
\newblock Functional sparse k-means clustering.
\newblock {\em Thesis, , Politecnico di Milano}.

\bibitem[Franti and Sieranoja, 2019]{FrantiSieranoja2019}
Franti, P. and Sieranoja, S. (2019).
\newblock How much can k-means be improved by using better initialization and
  repeats?
\newblock {\em Pattern Recognition}, 93:95 -- 112.

\bibitem[Garc{\'i}a et~al., 2015]{GarciaEtAl2015}
Garc{\'i}a, M. L.~L., Garc{\'i}a-Rodenas, R., and G{\'o}mez, A.~G. (2015).
\newblock K-means algorithms for functional data.
\newblock {\em NEUROCOMPUTING}, 151:231--245.

\bibitem[Hartigan, 1975]{Hartigan1975}
Hartigan, J.~A. (1975).
\newblock {\em Clustering algorithms}.
\newblock Wiley.

\bibitem[Hartigan and Wong, 1979]{HartiganWong1979}
Hartigan, J.~A. and Wong, M.~A. (1979).
\newblock Algorithm as 136: A k-means clustering algorithm.
\newblock {\em Journal of the Royal Statistical Society. Series C (Applied
  Statistics)}, 28(1):100--108.

\bibitem[Hu et~al., 2019]{Hu19}
Hu, G., Kaur, M., Hewage, K., and Sadiq, R. (2019).
\newblock Fuzzy clustering analysis of hydraulic fracturing additives for
  environmental and human health risk mitigation.
\newblock {\em Clearn Technologies and Environmental Policy}, 21:39--53.

\bibitem[Ieva et~al., 2013]{IevaEtAl2013}
Ieva, F., Paganoni, A.~M., Pigoli, D., and Vitelli, V. (2013).
\newblock Multivariate functional clustering for the morphological analysis of
  electrocardiograph curves.
\newblock {\em Journal of the Royal Statistical Society: Series C (Applied
  Statistics)}, 62(3):401--418.

\bibitem[Jim\'enez-Murcia et~al., 2019]{Murcia19}
Jim\'enez-Murcia, S., Granero, R., Fernndez-Aranda, F., Stinchfield, R.,
  Tremblay, J., Steward, T., Mestre-Bach, G., Lozano-Madrid, M., Mena-Moreno,
  T., Mallorqu\'i-Bagu\', N., Perales, J.~C., Navas, J.~F., Soriano-Mas, C.,
  Aymam\'i, N., G\'omez-Pea, M., Ag\"uera, Z., del Pino-Guti\'errez, A.,
  Mart\'in-Romera, V., and Mench\'on, J.~M. (2019).
\newblock Phenotypes in gambling disorder using sociodemographic and clinical
  clustering analysis: an unidentified new subtype?
\newblock {\em Front Psychiatry}, 10:173.

\bibitem[Karlin, 1973]{karlin1973}
Karlin, S. (1973).
\newblock Some variational problems on certain sobolev spaces and perfect
  splines.
\newblock {\em Bull. Amer. Math. Soc.}, 79(1):124--128.

\bibitem[Kooperberg and Stone, 1992]{KooperbergStone1992}
Kooperberg, C. and Stone, C.~J. (1992).
\newblock Logspline density estimation for censored data.
\newblock {\em Journal of Computational and Graphical Statistics},
  1(4):301--328.

\bibitem[Lachout et~al., 2005]{LachoutEtAl2005}
Lachout, P., Liebscher, E., and Vogel, S. (2005).
\newblock Strong convergence of estimators as $\epsilon$n-minimisers of
  optimisation problemsof optimisation problems.
\newblock {\em Annals of the Institute of Statistical Mathematics},
  57(2):291--313.

\bibitem[Lemaire, 1983]{Lemaire1983}
Lemaire, J. (1983).
\newblock Proprietes asymptotiques en classification.
\newblock {\em Statistiques et analyse des donnees}, 8:41--58.

\bibitem[Lindemann and LaValle, 2009]{LindemannLaValle2009}
Lindemann, S.~R. and LaValle, S.~M. (2009).
\newblock Simple and efficient algorithms for computing smooth, collision-free
  feedback laws over given cell decompositions.
\newblock {\em The International Journal of Robotics Research}, 28(5):600--621.

\bibitem[Mallat, 2008]{Mallat2008}
Mallat, S. (2008).
\newblock {\em A Wavelet Tour of Signal Processing, Third Edition: The Sparse
  Way}.
\newblock Academic Press, Inc., USA, 3rd edition.

\bibitem[Martino et~al., 2019]{MartinoEtAl2019}
Martino, A., Ghiglietti, A., Ieva, F., and Paganoni, A.~M. (2019).
\newblock A k-means procedure based on a mahalanobis type distance for
  clustering multivariate functional data.
\newblock {\em Statistical Methods {\&} Applications}, 28(2):301--322.

\bibitem[Pena et~al., 1999]{PenaEtAl1999}
Pena, J., Lozano, J., and Larranaga, P. (1999).
\newblock An empirical comparison of four initialization methods for the
  k-means algorithm.
\newblock {\em Pattern Recognition Letters}, 20(10):1027 -- 1040.

\bibitem[Reif, 1997]{Reif1997}
Reif, U. (1997).
\newblock Uniform b-spline approximation in sobolev spaces.
\newblock {\em Numerical Algorithms}, 15(1):1--14.

\bibitem[Tarpey and Kinateder, 2003]{TarpeyKinateder2003}
Tarpey, T. and Kinateder, K. K.~J. (2003).
\newblock Clustering functional data.
\newblock {\em Journal of Classification}, 20(1):093--114.

\bibitem[Tokushige et~al., 2007]{TokushigeEtAl2007}
Tokushige, S., Yadohisa, H., and Inada, K. (2007).
\newblock Crisp and fuzzy k-means clustering algorithms for multivariate
  functional data.
\newblock {\em Computational Statistics}, 22(1):1--16.

\bibitem[Udler et~al., 2018]{Udler18}
Udler, M.~S., Kim, J., von Grotthuss, M., Bons-Guarch, S., Cole, J.~B., Chiou,
  J., Anderson, C.~D., Boehnke, M., Laakso, M., Atzmon, G., Glaser,
  B.~Mercader, J.~M., Gaulton, K., Flannick, J., Getz, G., and Florez, J.~C.
  (2018).
\newblock Type 2 diabetes genetic loci informed by multi-trait associations
  point to disease mechanisms and subtypes: A soft clustering analysis.
\newblock {\em {PLOS} Medicine}.

\bibitem[Unser, 1997]{Unser1997}
Unser, M.~A. (1997).
\newblock {Ten good reasons for using spline wavelets}.
\newblock In Aldroubi, A., Laine, A.~F., and Unser, M.~A., editors, {\em
  Wavelet Applications in Signal and Image Processing V}, volume 3169, pages
  422 -- 431. International Society for Optics and Photonics, SPIE.

\bibitem[Wang et~al., 2014]{WangEtAl2014}
Wang, G., Lin, N., and Zhang, B. (2014).
\newblock {Functional k-means inverse regression}.
\newblock {\em Computational Statistics \& Data Analysis}, 70(C):172--182.

\bibitem[{Ward Jr.}, 1963]{Ward1963}
{Ward Jr.}, J.~H. (1963).
\newblock Hierarchical grouping to optimize an objective function.
\newblock {\em Journal of the American Statistical Association},
  58(301):236--244.

\bibitem[Wedel and Kamakura, 1999]{Wedel99}
Wedel, M. and Kamakura, W. (1999).
\newblock {\em Market segmentation: conceptual and methodological foundations}.
\newblock Springer Science \& Business Media, 2 edition.

\bibitem[Yamamoto, 2012]{Yamamoto2012}
Yamamoto, M. (2012).
\newblock Clustering of functional data in a low-dimensional subspace.
\newblock {\em Advances in Data Analysis and Classification}, 6(3):219--247.

\bibitem[Yamamoto and Terada, 2014]{YamamotoTerada2014}
Yamamoto, M. and Terada, Y. (2014).
\newblock Functional factorial k-means analysis.
\newblock {\em Computational Statistics and Data Analysis}, 79:133--148.

\bibitem[Zhang et~al., 1995]{ZhangEtAl1995}
Zhang, X., Begleiter, H., Porjesz, B., Wang, W., and Litke, A. (1995).
\newblock Event related potentials during object recognition tasks.
\newblock {\em Brain Research Bulletin}, 38(6):531 -- 538.

\end{thebibliography}


%
%



\end{document}